\documentclass[num-refs]{wiley-article}

\usepackage{siunitx}
\usepackage{amsmath}
\usepackage{tcolorbox}
\usepackage{graphicx}
\usepackage{subfig}
\usepackage{float}
\usepackage{setspace}
\usepackage{mathtools,xparse}
\usepackage{bbm}
\usepackage{textcomp}%
\usepackage{hyperref} 
\hypersetup{hidelinks,colorlinks,breaklinks=true,urlcolor=black,citecolor=blue,linkcolor=blue,bookmarksopen=false,pdftitle={Title},pdfauthor={Author}}

\DeclarePairedDelimiter{\abs}{\lvert}{\rvert}
\DeclarePairedDelimiter{\norm}{\lVert}{\rVert}

\papertype{Original Article}
\paperfield{Financial Mathematics}

\title{The CMMV Pricing Model in Practice}

\abbrevs{CMMV, Continuous Martingale of Maximal Variation; SS, Sticky Strike.}

\author[*]{Bernard DE MEYER}
\author[*]{Moussa DABO}

\affil[*]{PSE/CES -- Université Paris 1 - Panthéon Sorbonne}

\corraddress{Bernard DE MEYER, 106-112 Boulevard de l'Hôpital, 75647 Paris Cedex 13, France.}
\corremail{demeyer@univ-paris1.fr}

\fundinginfo{CES - ED 465}

\runningauthor{DE MEYER \& DABO}

\begin{document}

\maketitle

\begin{abstract}
Mainstream financial econometrics methods are based on models well tuned to replicate price dynamics, but with little to no economic justification. In particular, the randomness in these models is assumed to result from a combination of exogenous factors. In this paper, we present a model originating from game theory, whose corresponding price dynamics are a direct consequence of the information asymmetry between private and institutional investors. This model, namely the CMMV pricing model, is therefore rooted in market microstructure. The pricing methods derived from it also appear to fit very well historical price data. Indeed, as evidenced in the last section of the paper, the CMMV model does a very good job predicting option prices from readily available data. It also enables to recover the dynamic of the volatility surface.

\keywords{Game Theory --- Information asymmetry --- CMMV --- Option pricing}
\end{abstract}

\section*{Introduction}
Financial markets inherently exhibit instances of information asymmetry. Indeed, institutional investors have premium access to relevant financial information and are better equipped to analyze it. They generally have a sizable staff group devoted to analyzing economic conjunctures and the state of publicly listed companies. This provides them with an edge regarding the prediction of the long-term evolution of stock prices for example. In the short-term also, there are information asymmetries: investment banks generally operate as intermediaries in exchanges between other market participants. When they receive a large order to execute, this gives them some private information on the short-term evolution of the corresponding price. The optimal way to execute such orders on the market is the subject of the flourishing literature stemming from \cite{almgren2001optimal}, although not from an equilibrium perspective.

As a result of their informational advantage, financial institutions' moves are closely monitored by the other agents, in order to infer signals on the evolution of prices. The resulting trading setting can be properly modeled as a repeated exchange game with incomplete information. This approach allows to analyze the optimal use that can be made of private information. De Meyer and Saley (2003)\cite{meyer2003strategic}, De Meyer (2010)\cite{de2010price}, Gensbittel (2010)\cite{gensbittel2010analyse} and De Meyer and Fournier (2017)\cite{de2017price} show that in this game theoretic framework, the strategic use of information by the better-informed agents leads to very particular price dynamics. These are referred to as Continuous Martingales of Maximal Variation (CMMV hereafter).

The CMMV class of dynamics appears very robust in the sense that it is independent of the particular trading mechanism: it is the remote consequence of a central limit theorem. In a concise review of the aforementioned papers, we outline the market microstructure roots of the CMMV class. This will bring us to the CMMV hypothesis that paves the way for 2 simple pricing models. Before presenting these (in Section 3), we give an exhaustive account of the CMMV class in Section 2. Last but not least, the pricing models fit very accurately historical price data. Evidence of this fact is given in the last section of the paper.

\section{Literature review}

De Meyer and Moussa Saley (2003)\cite{meyer2003strategic} model the market as a game of incomplete information between two risk neutral market makers. More specifically, the informed agent (Player 1) receives some private information which will influence the value $L$ of the risky asset at a final date $T$. Player 2 only knows the probability distribution $\mu$ of $L$. Before that final date, there is a large number $n$ of trading periods. In each period, both players have to post prices. As a usual rule for market makers, the price they post is only a commitment to buy or sell a limited amount of shares at that price (there is no bid-ask spread allowed). In this model, the market maker with the lowest price gives one unit of the risky asset to her opponent in counterpart of $L_t$ units of the numéraire asset, $L_t$ being the price posted by her opponent. In this setup, one can easily understand that the informed agent has to find a trade-off between maximizing the current stage pay-off and staying with unrevealed information for the next stages. As shown in \cite{meyer2003strategic}, the only way for Player 1 to realize this trade-off is to use \textit{mixed} strategies, which amounts to introducing random noises on her actions. When the strategies are optimal and $n$ is tending to infinity, those random noises, introduced day after day for strategic reasons aggregate in a Brownian motion, thence providing an endogenous justification for the presence of random processes in financial models. 

De Meyer generalizes this result in \cite{de2010price} and further characterizes the limiting price process. In this paper, the market is modeled as a repeated game of incomplete information between one informed agent and the remaining part of the market. The second player in this setting is thus an aggregate of individual agents. Both players are considered as risk neutral. An action for player 2 is in fact a profile of actions of all its constituent agents. The action set of player 2 is thus very complicated, and one has to introduce the notion of abstract trading mechanism to model this situation: such a mechanism is a game where the two players select their actions in abstract spaces $I$ and $J$ and whose outcome is a transfer of shares and numéraire from Player 2 to Player 1. To model realistic markets, this trading mechanism has to satisfy a set of 5 hypotheses. When those fairly general hypotheses are satisfied, De Meyer (2010)\cite{de2010price} shows that as $n \rightarrow \infty$ , the price process in the $n$-times repeated game converges in distribution to a CMMV, regardless of the specific trading mechanism.

The previous papers consider only one risky asset, that is traded in exchange of a numéraire good, in modeling the market. However, Gensbittel (2010)\cite{gensbittel2010analyse} confirms their findings in a multi-asset model. In details, if the players exchange simultaneously one underlying asset and a bunch of monotonic European derivatives as call or put options with the same expiry date, then the prices of all assets are asymptotically CMMVs. 

In addition, De Meyer and Fournier (2017)\cite{de2017price} introduced the risk aversion of private investors in the model by representing the uninformed part of the market as a risk averse agent. The resulting game is no longer zero-sum. Though their first finding is that the price process is compatible with the usual no arbitrage theory, in the sense that there exists a unique Martingale Equivalent Measure (MEM hereafter) under which the price process becomes a martingale. They further prove that, as $n \rightarrow \infty$, the price process under the MEM converges to a CMMV. 

In all these models, the numéraire yields no interest. If there were a positive interest rate, the same conclusion could be derived for the discounted price process. This last observation justifies what we refer to as the \textcolor{red}{\textbf{CMMV hypothesis}}: \textbf{\textit{Under the MEM, the discounted stock prices are CMMVs.}}

\section{The CMMV class}

\subsection{Definition}

A CMMV on a time interval $[0, T]$ is a martingale $(\Pi_t)_{t \in [0, T]}$ that can be written as: $\forall t, \Pi_t = f(B_t,t)$, 
where $B$ is a standard Brownian motion and $f : \mathbb{R} \times [0,T] \rightarrow \mathbb{R}$ is increasing with respect to its first argument. 

Noticeably, the function $f$ also has to satisfy the (time-reversed) heat equation: $$\frac{\partial}{\partial t} f_t(x) + \frac{1}{2} \frac{\partial^2}{\partial x^2}f_t(x) = 0 $$  for $(\Pi_t)_{t \in [0, T]}$ to be a martingale. As in the above equation, $f(x, t)$ will be denoted $f_t(x)$, in the rest of the paper. 

It turns out that the CMMV class encompasses 2 of the most prominent (at least historically) financial pricing models, namely Bachelier: $f_t(x) = a x + b$ and Black-Scholes-Merton: $f_t(x) = a e^{\sigma x - \frac{\sigma^2}{2} t}$, for some $a > 0$. Furthermore, $x \mapsto f_t(x)$ admits an inverse, being strictly increasing. So there is some function $\psi_t$ such that $B_t = \psi_t(\Pi_t)$. The CMMV model is therefore a sub-class of local volatility models since $d\Pi_t = \frac{\partial}{\partial x} f_t(B_t) d B_t = \nu(\Pi_t, t) d B_t$, $\nu: (y, t) \mapsto \frac{\partial}{\partial x} f_t(\psi_t(y))$ being the volatility function (see \cite{dupire1994pricing}).

To wrap up the definition, we shed some light on the CMMV terminology. It actually finds its origin in the informed player's optimization problem. Indeed, in a risk neutral setting, while acting upon her private information, the informed player will change the perception of the uninformed player, thereby modifying the expected price $L_q = E[L|\mathcal{F}_q]$ (with $\mathcal{F}_q$ denoting the public information at date $q$). In a game of lenght $n$, Player 1 may select any martingale $(L_1, ..., L_n, L)$, provided $L$ is $\mu-$distributed. Roughly speaking, the stage $q$ pay-off will be proportional to $|L_{q+1} - L_{q}|$ and Player 1 will thus have to maximize the $L^1-$variation:    
$$\sum_{q=0}^{n-1}  \norm{L_{q+1} - L_q}_{L^1}$$
on the class of martingales $(L_1, ..., L_n, L)$, with $L$ being $\mu-$ distributed. The optimal martingale $L^n$ in the problem of length $n$ and more precisely the continuous time representation of $L^n$, the process $L^n_{\lfloor nt \rfloor}$ (${\lfloor x \rfloor}$ denoting the integer part of x), is shown in \cite{de2010price} to converge to a CMMV on the $[0, 1]$ time interval. This justifies the CMMV terminology.

\subsection{Basic properties of a CMMV}

We present in this sub-section some properties of a CMMV that will be useful in the next sections. The reader may refer to \cite{de2010price} for a more detailed presentation. 

Firstly, we remark that the function $f$ associated with the CMMV satisfies very interesting regularity conditions. Indeed, by means of the Markov property of the Brownian motion, we have $\forall t \in [0,T]$
\begin{equation*}
\begin{split}
\Pi_t & =  E[\Pi_T|\mathcal{F}_t] = E[f_T(B_T)|\mathcal{F}_t] \\ & = E[f_T(B_t + (B_T - B_t))|\mathcal{F}_t] = f_t(B_t) 
\end{split}
\end{equation*} 
where ($\mathcal{F}_t$) is the Brownian filtration and this implies:
\begin{equation}
f_t(x) = E[f_T(x + \sqrt{T-t}Z)]
\label{eq:f(B,t)}
\end{equation} 
with $Z$ \texttildelow $\mathcal{N}(0,1)$. 

In other words, \eqref{eq:f(B,t)} indicates that, $\forall t \leq T$, 
\begin{equation}
f_t(x)= \int_{-\infty}^{\infty}f_{T}(y)h_{T-t}(x-y)dy =f_T * h_{T-t}(x)
\label{Conv}
\end{equation}
$f$ is therefore the convolution of $f_T$ (the function at the final date) with the  $\mathcal{N}(0,T-t)$ - density function $h_{T-t}$. Thanks to the smoothness of the normal kernel, if $f_T(\sqrt{T - t}Z) \in L^2$ then the integral in (\ref{Conv}) converges $\forall x \in \mathbb{C}$. The resulting function is entire (holomorphic on the whole complex plane). $f$ is, therefore, $\mathcal{C}^\infty$ on $\mathbb{R} \times [0,T[$. We also have : $f_t^{(k)}= f_T * h^{(k)}_{T-t}$, with $f^{(k)}_t$ (respectively $h^{(k)}_{T-t}$) denoting the $k$-th order partial derivative of $f_t$ (respectively $h_{T-t}$).  

By applying Itô's lemma on the CMMV, we also have:

\begin{equation}
d\Pi_t = f^{(1)}_{t}(B_t)dB_t 
\label{eq:Dpit}
\end{equation}
There is no drift term since by definition $(\Pi_t)_{t\in[0,T]}$ is a martingale. The convolutive form of $f_t$ also indicates that, if $f_T$ is $k$ times differentiable:  

\begin{equation}
f^{(k)}_t = f^{(k)}_T * h_{T-t} 
\label{fkt}
\end{equation}
As a consequence, $f^{(k)}_t(B_t)$ is also a martingale since $f^{(k)}_t(x) = E[f^{(k)}_T(x + \sqrt{T-t}Z)]$ and thus $$f^{(k)}_t(B_t) = E[f_T^{(k)}(B_t + (B_T - B_t))|\mathcal{F}_t] = E[f^{(k)}_t(B_t + \sqrt{T-t}Z)]$$

Another important property is that, for a given probability distribution $\mu$ satisfying $\int_{-\infty}^{\infty} \abs x d\mu(x) < \infty$ , there exists a unique CMMV $\Pi^{\mu,T}$ such that $\Pi^{\mu,T}_T$ is $\mu-$distributed. In fact, there exists a unique right-continuous increasing function $f^{\mu} : \mathbb{R} \rightarrow \mathbb{R}$ that satisfies $f^{\mu}(Z) $ \texttildelow $\mu$ when $Z$ \texttildelow $\mathcal{N}(0,T)$. $f^{\mu}$ can be expressed as: 
\begin{equation}
f^{\mu}(x) = F^{-1}_{\mu}(F_{\mathcal{N}(0,T)}(x))
\label{eq:fmu}	
\end{equation} 
where $F_d$ denotes the cumulative distribution function of $d$ and $F^{-1}_{\mu}$ is defined by $F^{-1}_{\mu}(x) := \inf\{y: F_{\mu}(y) > x\}$.

We further have $\forall x \in \mathbb{R}, f_T(x) = f^{\mu}(x)$ and $f$ can be recovered as previously, using 
\begin{equation}
f_t(x) = \int_{-\infty}^{\infty}f^{\mu}(y)h_{T-t}(x - y)dy
\label{eq:conv}	 
\end{equation}
The fact that $f$ satisfies the time reversed heat equation also implies that $f$ can be expressed as a  series of Hermite functions. Indeed, defining the scalar product $\langle.,.\rangle_T$ of 2 functions as :  $$ \langle g_1 , g_2\rangle_T = \int_{-\infty}^{\infty} g_1(x) h_T(x) g_2(x)dx$$ we have that Hermite polynomials $(H_n)_{n \in \mathbb{N}}$ are orthogonal for the scalar product: $\langle.,.\rangle_1$. They are given by the generating formula: $H_n(x) = (-1)^n e^{\frac{x^2}{2}} \frac{d^n}{d x^n}(e^{-\frac{x^2}{2}})$ (see \cite{beck2004objectif}).

Therefore considering the functions: $$\phi_n : (x,t) \mapsto H_{n}(\frac{x}{\sqrt{t}}) (\sqrt{t})^n, \ \forall n \in \mathbb{N}$$
we obtain a sequence $(\phi_n(.,t))_{n \in \mathbb{N}}$ of orthogonal polynomials for each weight function $h_t$.  

By Theorem $3.49$ in \cite{beck2004objectif}, $(\phi_n(.,t))_{n \in \mathbb{N}}$ is an Hilbert basis for the aforementioned inner product space, since $\int_{-\infty}^{\infty} e^{\abs{x}}  h_T(x) dx = e^{\frac{T}{2}} < \infty$. Therefore if $f_T \in L^2(h_T)$, and setting $\epsilon_n(x)$ as:

\begin{equation}
\epsilon_n(x) = \abs{f_T(x) - \sum_{k = 0}^{n} \alpha_k \phi_k(x,T)}
\label{devfT}
\end{equation} 
where $\alpha_k = \frac{\langle \phi_k, f_T\rangle}{\langle \phi_k, \phi_k\rangle}$, we have $\| \epsilon_n \|_{L^2(h_t)} \rightarrow 0$. It follows therefore that:

\begin{equation*}
	\begin{split}
		\abs{f_0(x) - \sum_{k = 0}^{n} \alpha_k x^k} & \leq \epsilon_n * h_T (x)   \\ & = \int_{-\infty}^{\infty}  \epsilon_n(z) h_T(x - z) dz
	 \\ & = e^{- \frac{x^2}{2}} \int_{-\infty}^{\infty} \epsilon_n( z) e^{x z} h_T(z)  dz \\ & = e^{- \frac{x^2}{2}} \langle \epsilon_n (z), e^{x z}\rangle_T \\ & \leq e^{- \frac{x^2}{2}} \| \epsilon_n (z) \|_{L^2(h_T)} \| e^{x z} \|_{L^2(h_T)} \rightarrow 0 \ \text{as} \ n \rightarrow \infty
	\end{split}
\end{equation*}

We have thus proved that $f_0(x) = \sum_{k = 0}^{\infty} \alpha_k x^k$. Since $f_0$ is an entire function, this indicates that $\forall k \in \mathbb{N}, \alpha_k = \frac{f_0^{(k)}(0)}{k!}$. It results from the above that knowing all the derivatives at $t = 0$, allows to recover the function $f_t$ at any other date.

\section{Pricing with the CMMV model}

It turns out that the \textit{CMMV Hypothesis} makes it possible to compute derivative pricing formulae and hedging strategies. Due to its above-mentioned market microstructure foundation, the methods derived from it are potentially more robust than mainstream methods based on rules of thumb. The point is that knowing the CMMV function $f$ of the prices under the MEM $Q$ gives a complete knowledge of the price dynamics under $Q$ (unless otherwise stated, all prices are discounted). And this leads to explicit pricing formulas for derivatives. For example, under the MEM $Q$, the price process of a European call option $(C_{K,t}^T)_{t \in [0,T]}$, whose underlying asset has price process $(S_t)_{t \in [0,T]}$, is given by: 
\begin{equation}
\begin{split}
C_{K,t}^T & = E[(S_T - K)^+ | \mathcal{F}_t] \\ & = E[(f_T(B_t + (B_T - B_t)) - K)^+ | \mathcal{F}_t] \\ & = g^{K, T}_t(B_t) 
\end{split}
\label{eq:exp}
\end{equation}
where $(x)^+ = Max\{x,0\}$ and $g^{K, T}_t(x) = E[(f_T(x + \sqrt{T-t} Z) - K)^+]$, $Z \texttildelow \mathcal{N}(0, 1)$. In this last formula, $B_t$ can be recovered from $S_t$ since $S_t = f_t(B_t)$ and $f_t$ is increasing. 

Note that, since $f_t$ is increasing, so is $g^{K, T}_t$. It follows, therefore, that $C^T_{K,t} $ is also a CMMV under the MEM. This is a consistency argument for the CMMV pricing model. Indeed, if a family of dynamics is supposed to govern the price processes of all stocks on the market, it also should fit the price processes of their derivatives. Many popular models, like the Black-Scholes-Merton model, fail to meet this requirement. 

Knowing the CMMV function $f$, one could also price any complicated derivative using for instance Monte-Carlo methods. This is the motivation for knowing the function associated with the CMMV. In the following paragraphs, we present methods of recovering this function from historical data. Firstly, we consider the case in which historical prices are observed in continuous time, that is one sample path of the price process is observed on a time-interval  $[0,t_1]$. All we know in this case is that under an equivalent probability measure, this price process is a CMMV. Still, the function can (theoretically) be fully recovered with no statistical error, as shown in the next sub-section. However, this result is purely theoretical and cannot be used to recover $f$ in practice because prices are only available in discrete time. And in the discrete case, it is impossible to compute the quadratic variation process and its infinite  sequence of derivatives (required to recover $f$ in the continuous setting). 

In fact, the problem of estimating $f$ based on discrete time observation of the price trajectory is statistical. Though, it involves the unknown density of the MEM which can be arbitrary. The problem is as difficult to solve as would be the one of estimating the variance of a random variable based on a single realization and the sole knowledge that the random variable has a density. It is however possible to go around the problem of the unknown density of the MEM. We provide in the subsection 3.2.1 a method of recovering the function $f$ from the observation at date $t=0$ of European call option prices (with a single maturity) for a large number of strikes. There is also a method presented in subsection 3.2.2 that requires only the observation of the price processes of the asset and one single European call option. 

\subsection{Continuous-time observation of $(S_t)$}

Suppose we observe the historical prices $(S_t)_{t\in[0,t_1]}$ of an asset in continuous time and full precision on the interval $[0,t_1] \ (t_1 < T)$. Is it possible to find the function $f : \mathbb{R}$x$[0,T] \rightarrow \mathbb{R}$, increasing in its first argument such that:  
\begin{itemize}
	\item[-] $S_t = f_t(B_t), \forall t \in [0,T] $ and
	\item[-] Under the MEM $Q$, B is a Brownian motion
\end{itemize}
This question does have a positive answer. Indeed, we firstly observe that by \eqref{eq:Dpit},  $dS_t = f^{(1)}_t(B_t)dB_t$. We can next define $S^n_t := f^{(n)}_t(B_t)$. As explained in \eqref{fkt}, $(S^n_t)_{t \in [0,t_1]}$ is a martingale and making use of Itô's formula, one gets: 

\begin{equation}
dS^{n-1}_t = f^{(n)}_t(B_t)dB_t = S_t^n dB_t
\label{DSn-1}
\end{equation}
We next claim that it is possible to recover the path of all $(S^n_t)_{t \in [0,t_1]} \forall n$. For now, note that from the path of $(S_t)_{t \in [0,t_1]}$ one can compute\footnote{\label{foo}Given the 2 paths of semimartingales $X$ and $Y$, it is indeed possible to recover with probability 1 the path of $\langle X, Y \rangle$. The covariation $\sum (X_{t_{i+1}} - X_{t_i})(Y_{t_{i+1}} - Y_{t_i})$ on a discrete subdivision of $[0, T]$ will converge in probability to $\langle X, Y \rangle$ as the mesh tends to 0\cite{revuz1999continuous}. It is then possible to find a subsequence of subdivisions that ensures the almost-sure convergence. So with probability 1, it is possible to compute $\langle X, Y \rangle_t$ for all rational $t \in [0, t_1]$. By continuity, we get then the whole path.}  the process $\left\langle S, S\right\rangle$:

\begin{equation}
\left\langle S, S\right\rangle_t = \int_{0}^{t}\left( S_s^1\right)^2d\left\langle B, B\right\rangle_s =  \int_{0}^{t}\left( S_s^1\right)^2ds
\end{equation}
Differentiating with respect to $t$ one gets:

\begin{equation}
(S_t^1)^2 = \frac{d}{dt}\left\langle S, S\right\rangle_t
\end{equation}
since $f$ is strictly increasing with respect to its first argument, we have $f^{(1)}_t(x) > 0, \ \forall x \in \mathbb{R}$. Hence, $(S^1_t)$ is strictly positive and thus:

\begin{equation}
S_t^1 = \sqrt{\frac{d}{dt}\left\langle S, S\right\rangle_t}
\label{St1}
\end{equation}
It is then possible to recover\footnote{The Riemann-Stieltjes sums $\sum \frac{1}{S_{t_i}} (S_{t_{i+1}} - S_{t_i})$ approximating the Itô integral converge in probability as the mesh of the subdivision goes to 0. Consequently, by selecting a appropriate subsequence, we get almost-sure convergence on all rational $t \in [0, t_1]$. And this can be extended to all $t \in [0, t_1]$.} the path of the Brownian motion $B$ from:

\begin{equation}
B_t = \int_{0}^{t} \frac{1}{S_s^1} dS_s
\end{equation}
Knowing $(S^{n-1}_t)_{t \in [0, t_1]}$ and the Brownian trajectory, one can compute\textsuperscript{\ref{foo}} the process $\left\langle S^{n-1}, B \right\rangle$. Making use of \eqref{DSn-1}, we have:  
\begin{equation}
\left\langle S^{n-1}, B \right\rangle_t = \int_{0}^{t}S^n_sds
\end{equation} 
And thus:
\begin{equation}
S^n_t = \frac{d}{dt}\left\langle S^{n-1}, B \right\rangle_t
\end{equation}
We get in this way all the $S^n_t,  \forall n$ using a recursion. In particular, we obtain $f^{(n)}_0(0)$, $\forall n \in \mathbb{N}$ since $f^{(n)}_0(0) =  S^n_0$. According to the last remark of section 2, we obtain the whole function $f_t$ since $f_t(x) = \sum_{n = 0}^{\infty} \frac{f_0^{(n)}(0)}{n!} \phi_n(x, t)$.

The previous discussion indicates that all functions $f$ characterizing the dynamics of a CMMV can be recovered from the observation of a single generic trajectory. By generic we mean selected in a set of trajectories that has probability $1$ under the MEM. Since the MEM is equivalent to the historical probability, this is also a set of probability $1$ under the historical measure.

In this setting, the problem of estimating the CMMV function $f$ is deterministic rather than statistical: indeed, if $\Pi_f$ and $\Pi_g$ are 2 CMMVs corresponding to 2 different functions $f_t(.)$ and $g_t(.)$ (i.e.  $\Pi_{f,t} = f_t(B_t)$, and  $\Pi_{g,t} = g_t(B_t)$), then the measures $\mu_f$ and $\mu_g$ they induce on the set $\mathcal{C}([0,T]) $ of continuous  trajectories are mutually singular.

In the next subsection, we treat the more realistic case of observations in discrete time.

\subsection{Discrete-time observations}

In the discrete case, the problem can be formulated as follows: given the historical observations $(S_t)_{t\in\{t_1,t_2, ...,t_n\}}$ with $0=t_1<t_2<...<t_n<T$, is it possible to find the function $f : \mathbb{R}$x$[0,T] \rightarrow \mathbb{R}$, increasing in its first argument such that:  
\begin{itemize}
	\item[-] $S_t = f_t(B_t), \forall t \in [0,T] $ and
	\item[-] Under the MEM $Q$, B is a Brownian motion 
\end{itemize}
There is no satisfactory answer to this question because the above problem is statistical and non-parametric in a double sense: the observation space is $\mathbb{R}^n$ and if $(\Omega, \mathcal{A}, Q)$ is the probability space where the Brownian motion is defined, the historical probability $P$ is determined by the density $y = \frac{dP}{dQ}$, which is arbitrary. Hence, the parameter vector for the statistical model is $\theta = (y, f_T)$, with both $y$ and $f_T$ belonging to infinite dimensional spaces. 

We would like to estimate $f_T$ based on the observation of a random vector $X \in \mathbb{R}^n$ where, $\forall t \in \{t_1,t_2, ...,t_n\}$, $X_t = f_t(B_t)$. The probability distribution $P_{\theta}$ of $X$ is the law of $(f_t(B_t(\omega)))_{t\in\{t_1,t_2, ...,t_n\}}$ when the underlying probability space $\Omega$ is endowed with the probability $y.dQ$. For each increasing function $\tilde{f}_T$, one could compute all functions $\tilde{f}_t$ by (\ref{eq:conv}) and solve for $x_t$ in the equation $\tilde{f}_t(x_t) = X_t$. To estimate the true $f_T$, one would then select the $\tilde{f}_T$ that makes $(x_t)$ compatible with the law of $(B_t)$ under $y.dQ$. The unknown density $y$ makes it clearly impossible to carry this out. Therefore, it is fair to say there exists no satisfactory estimator of the parameter vector, due to the unknown density. 

In the next 2 sub-sections, we propose alternative ways of approximating the function associated with the CMMV, that need no additional assumptions. They will require some additional data though.

\subsubsection{Observation of $(C^K_t)$ for multiple strikes at date $t=0$ [M1]}

For some securities (especially indexes), there is a considerable number of derivatives traded at a high frequency. And it turns out that, for a stock with many options, one can approximate the CMMV dynamic from data on a single trading period.

More specifically, consider that the price of the stock is observed along with the prices of a large number of European call options on it, at an initial date $t=0$. As we will explain now, the statistical problem of estimating $f$ is in this case a deterministic one. Indeed, one has $C_{K,0}^T = E[(S_T - K)^+]$, where the expectation is taken with respect to $Q$ and the density $y$ is not involved in the expression. 

Differentiating the price of the call (under the MEM) with respect to $K$, we get:

\begin{equation}
\begin{split}
\frac{\partial}{\partial K}C_{K,0}^T & =  E[-\mathbbm{1}_{\{S_T > K\}}] = - Q[S_T > K] \\ & = F_{S_T}(K) - 1
\end{split}
\label{eq:ddck}
\end{equation} 
which gives the cumulative distribution function $F_{S_T}$ (hence the law $\mu$) of the random variable $S_T$ under $Q$. Now let $f_\mu : \mathbb{R} \rightarrow \mathbb{R}$ be the unique (right-continuous) increasing function such that $f_\mu(B_T)$ \texttildelow $\mu$. This is exactly the function at the final period: $\forall x \in \mathbb{R}, f_T(x) = f_\mu(x)$, which completely caracterises the CMMV function (see \ref{eq:conv}). In practice, we carry out the approximation in the following 3 steps:

\begin{enumerate}
	\item Using the dataset comprised of options prices and their corresponding strikes, estimate the derivatives $\frac{\partial}{\partial K}C_{K,0}^T$ by finite differences.   
	\item Bearing in mind (\ref{eq:fmu}), compute the value $\xi_K = F_{\mathcal{N}(0,T)}^{-1}(1 + \frac{\partial}{\partial K}C_{K,0}^T)$ for each strike $K$ in the dataset, $F_{\mathcal{N}(0,T)}$ being the cumulative distribution function of $\mathcal{N}(0,T)$.
	\item The pairs $(\xi_K, K)$ belong then to the graph of $f_\mu$ and the function $f_\mu$ is obtained by fitting this list to a polynomial model for instance.
\end{enumerate}  
Having determined $f_T = f_\mu$, we get the function $f_t$ at every date by (\ref{eq:conv}). As outlined at the beginning of  section 3, this leads to pricing formulas for any derivative. As will be evidenced in the results section, this pricing method fits real market data quite impressively. And it does so for long trading periods.

However, the data required to carry out this method (referred to as M1 hereafter) is rarely available. In fact, there is only one or a few options for most stocks. Therefore in the next sub-section, we present a method that requires, in addition to the price process of the stock, the price process of only one option.

\subsubsection{Discrete-time observation of $(S_t)$ and $(C_t)$ [M2]}

For most publicly traded securities, historical price time series are readily available. And in general so is price data on at least one option on them. In this sub-section, we show how to recover the CMMV function from longitudinal price data. Specifically, assume that in addition to the historical price process of the stock, we also observe the historical prices of a European call option, with strike $K$, for which it is the underlying. As for the previous case, the statistical problem becomes deterministic and the unknown density is no longer involved. It can in fact be reformulated as: given the historical observations $(S_t)_{t\in\{t_1,t_2, ...,t_n\}}$ and $(C_t)_{t\in\{t_1,t_2, ...,t_n\}}$, with $0=t_1<t_2<...<t_n<T$, is it possible to find the function $f : \mathbb{R}$x$[0,T] \rightarrow \mathbb{R}$, increasing in its first argument such that:  

\begin{enumerate}
	\item[-] $S_t = f_t(B_t), \forall t\in\{t_1,t_2, ...,t_n\}$,
	\item[-] $C_t = g_t^{K, T}(B_t), \forall t\in\{t_1,t_2, ...,t_n\}$, where $g^{K, T}_t(x) = E[(f_T(x + \sqrt{T - t}Z) - K)^+]$, $Z \texttildelow \mathcal{N}(0, 1)$
\end{enumerate}

The unobserved Brownian motion $(B_t)$ can now be considered as an intermediary variable. At each period $t$, there are 2 identities involving it. So we can solve for its corresponding expression as a function of the parameters from the first, and use this expression in computing the second. Because it is only intermediary, we do not have to care about its distribution under the MEM. More specifically, the method (M2 hereafter) we propose proceeds as follows:

\begin{enumerate}
	\item Fix the class of eligible functions (typically the class of increasing polynomials): $\mathcal{F}$.  
	\item Solve for $\hat{f} \in \mathcal{F}$ that minimizes (by an iterative process) the error function $err(\hat{f})$ defined by the following 4:
	\begin{enumerate}
		\item $\hat{f}_T$ being a polynomial, the function $\hat{f}$ can easily be computed in terms of Hermite functions.
		\item Equation $\hat{f}_{t_i}(x_{t_i}) = S_{t_i}$ can then be solved numerically to find $x_{t_i}$.
		\item
	 $\hat{g}^{K, T}_{t_i}(x_{t_i})= E[(\hat{f}_T(x_{t_i} + \sqrt{T - t}Z) - K)^+]$, $Z \texttildelow \mathcal{N}(0, 1)$ can then be evaluated numerically.
		\item Compute the error term $err(\hat{f}) = \Sigma_{i}(\hat{g}^{K, T}_{t_i}(x_{t_i}) - C_{t_i})^2$
	\end{enumerate}
\end{enumerate}
This method also fits market data very well, as will be evidenced in the next section.  

\section{Numerical results}
\subsection{The data}

The dataset with which we have completed our numerical analysis is a large panel of prices of options on the SPX index. It was purchased from \hyperlink{historicaloptiondata.com}{historicaloptiondata.com} in the spring of 2017 for academic research. In addition to the prices of the last closed deals (all in USD), it incorporates among other variables, the trade date, the traded volumes, the bid, ask and strike prices, the greeks... The period covered ranges from 1990 to 2017. It is of primary importance for our methods to have simultaneous prices of options and the underlying. This is why we do not use the prices from the dataset since we have no information regarding the time at which the transactions occurred. For options, we consider the price to be the average between the bid and the ask. Then at every quotedate, the underlying's price and the discounting factor are estimated via linear regression from the put-call parity identity.

\subsection{Optimization}

\subsubsection{Modelization specifics}

As mentioned before, we use the class $\mathcal{F}$ of increasing polynomials. Observing that a polynomial $f$ of (odd) degree $2 n + 1$ is increasing if and only if $f' = P^2 + Q^2$, with $deg(P) = n > deg(Q)$, the class $\mathcal{F}_{2n + 1}$ of increasing polynomials of degree $2n + 1$ can easily be parameterized with the coefficients of $P$ and $Q$ plus an integrating constant. Also, in theory, the accuracy of the approximation increases with the order of the polynomial, but a model with a great number of terms does incur the risk of overfitting the data. Hence, we include the order of the polynomial in the parameters and estimate it via cross-validation. More specifically we split our data into 2 subsets. The first is used to fit the polynomial for a number of degrees and then we pick the degree that minimizes the error on the test subset.  

Before applying our methods to market data, we tested them on simulated CMMVs with a few arbitrary functional forms for $f_T$. It appeared that for M2, classical optimization algorithms (Gradient Descent, Newton Methods and their variants) fail to properly minimize its corresponding error function. Hence, for the function approximation for M2, we opt for a stochastic evolutionary algorithm, the Covariance Matrix Adaptation Evolution Strategy (CMA-ES hereafter), which was quite efficient. Additionally, its convergence is not dependent on the starting point\cite{akimoto2012convergence}. In the next sub-section, we give a brief presentation of this algorithm.

\subsubsection{The CMA-ES algorithm}

This presentation focuses on practical aspects relevant to our problem and the intuition behind the strategy, rather than its theoretic mathematical foundations. The reader is referred to \cite{hansen2016cma} for a more detailed exposition and to \cite{hansen2006cma} and \cite{peng2010population} for a comparative review of the method.

As can be inferred from their denomination, Evolution Strategies (ES) are stochastic methods for numerical optimization that are based on biological evolution (Darwinism) principles. In essence, from an initial generation (set of search points), they select the ``fittest'' members (the fitness criteria being the value of the objective function) to be parents that through variation, give birth to the next generation. This ``natural selection'' mechanism produces fitter individuals at each generation (iteration) until reaching a point with satisfactory fitness level. Thence the main drawback of these methods is that they require at least as many function evaluations as there are offspring at every iteration, and the population size is positively correlated with the convergence. This is particularly problematic if the objective function is not ``cheap'' to evaluate, in which case the implementation can be extremely slow. But this is not an issue we face with our previously defined objective function. In fact, the CMA-ES compares favorably in terms of computational performance with the Newton-Raphson methods for quick-response functions\cite{peng2010population}.  Specifically, the algorithm involves a 3-stages procedure at every iteration:

\begin{itemize}
	\item Sampling of new candidate solutions based on already stored information about the distribution of the best points,
	\item Ordering the new sample based on the fitness criteria and 
	\item Updating the search distribution parameters based on this newly acquired information.
\end{itemize}
The following presents a simplidied pseudocode of the  CMA-ES algorithm.

\begin{tcolorbox}[adjusted title=CMA-ES Pseudocode ,halign=flush left, colframe=gray]
	Set the population size $N$ \\
	Initialize the parameters of the search distribution (state variables) $\ m, \sigma, C = I_N$  \\
	\textbf{While termination criteria not met}  \\
	Sample $N$ new candidate solutions using:  \\
	$v_k =$ SampleNormal $(mean=m, cov. matrix = \sigma^2 C)$  for $\ k = 1, ..., N$  \\
	Evaluate the objective function at these new points and rank them accordingly  \\
	Move mean to the $n$ ($n < N$) best solutions: $m =$ Average$(v_1, ... v_n)$  \\
	Update $C$ and $\sigma$ using the evolution paths and new sample	\\
	\textbf{Return $v_1$}
\end{tcolorbox}

In detail, the new search points at each iteration are generated by sampling a multivariate normal distribution\footnote{The justification for the use of the normal distribution is the fact that, of all distributions in $\mathbb{R}^n$, it has the largest entropy given all covariances.} by way of the following scheme:

\begin{equation}
v^{(g+1)} \sim m^{(g)} + \sigma^{(g)} \mathcal{N}(0, C^{(g)}), \quad 
\label{eq:sample}
\end{equation}
where:
\begin{itemize}
	\item $v^{(g+1)}$ is the group composing generation $g+1$
	\item $\mathcal{N}(0, C^{(g)})$ is the centered multivariate normal with covariance matrix $C^{(g)}$
	\item $m^{(g)}$ is the mean of the search distribution at generation $g$
	\item $\sigma^{(g)}$ is the overall standard deviation (step size) at generation $g$.
\end{itemize}

At every iteration, the adaptation of the model parameters is based on 2 principles. The first is a Maximum Likelihood principle: the mean is updated such that the likelihood of previously successful candidate solutions is maximized. This amounts to ``Selection'' and ``Recombination'': the new mean at generation $g+1$ is a weighted average of the $n$ best points from the sample $(v_1^{(g+1)}, ..., v_N^{(g+1)})$. The only criteria for selection is the ranking of these points based on their objective function value, which precludes restrictive assumptions on the function and search space\footnote{since we choose the class of polynomials, we restrict the search space to ensure monotonicity. This is done by adding a penalty to the objective function.}. Then, the covariance matrix is updated to increase the probability of previously successful search steps. Indeed, it is estimated by the empirical covariance matrix but instead of using the empirical mean of the new sample, the mean from the previous generation is used. This way, it estimates the variance of sampled steps $v_i^{(g+1)} - m^{(g)}$, instead of the distribution variance within the sampled points. In addition, it uses the weighted selection mechanism as in the computation of the mean and is given by:

\begin{equation}
C_n^{(g+1)} = \sum_{i=1}^n w_i (v_i^{(g+1)} - m^{(g)}) (v_i^{(g+1)} - m^{(g)})'
\end{equation}  
$(w_i)$ denoting the weights. 

Sampling from $C_n^{(g+1)}$ tends to reproduce successful search steps. However, it is an estimator and as such, its reliability depends on the sample size and increasing the number of offspring has a computational cost. This brings us to the second adaptation principle: storing and additionally using information from previous generations.  That is, information about the correlation of consecutive steps is recorded by 2 parameters (evolution paths) in order to increase the variance in favorable directions and prevent premature convergence. This aspect of the algorithm averts the need to use large population sizes, which is a typical issue associated with evolutionary algorithms. There are also a number of other intermediary parameters that are used to update the above-mentioned state variables at every iteration, but we skip to list them all because the objective here is just giving a general view about the model. \\

In view of its features, the CMA-ES is an attractive alternative to classical Quasi-Newton methods for non-linear optimization, especially for non-convex or rugged search landscapes. Actually, adaptation of the covariance matrix amounts to learning a second order model of the underlying objective function\footnote{The objective behind the adaptation of the covariance matrix is to approximate the contour lines of the objective function. On convex-quadratic functions this amounts to approximating the inverse Hessian matrix}, similar to the approximation of the inverse Hessian matrix in Quasi-Newton methods\cite{hansen2016cma}. It satisfies, in addition, very interesting invariance properties including (but not restricted to) translation invariance, invariance to strictly monotone transformations and to angle preserving transformations. Last but not least, all estimates of the parameters that appear in the basic sampling equation \eqref{eq:sample} satisfy stationarity and unbiasedness.

\subsection{Results and discussion}

This part is devoted to the presentation of the results we obtained from the application of M1 and M2 to real market data. The data cleaning phase was carried out using $R$, version $3.5.3$ and the optimization phase was implemented in Mathematica, version $11.0.1.0$. 

We present the results of our 2 methods and those obtained from the Sticky Strike (SS hereafter) pricing model. This enables us to make a comparative analysis of their predictive performance. For the purposes of this application, we chose options with expiry date January 20th 2017, their liquidity (most traded options in our dataset) being the determining selection criterion. The first options with this expiry date and strike prices ranging from 100 to 3500 in our dataset were quoted on the 10th of August 2015, the then (undiscounted) spot price of the index being 2102.60. The data we have on these options spans therefore a period of over 500 days. Though we focus only on the period from the 20th of July 2016 onwards (184 days to expiration). On that day, 96 call (and 96 put) options with our chosen expiry date were quoted. We make use of their prices to implement M1. For M2, we use the option with strike price 2100, which is quoted (and traded) on every trading day in the period considered. For this method, our training set consists therefore of 44 prices observed in the first 2 months, which is around 30\% of the total number of trading days. The remaining data (4 months worth of observations) will serve as test set.

The first important result we obtain is that our 2 methods, although based on different approaches, fit the CMMV function with very close third order polynomials. In fact, if we consider the interval stretching from the first to the 99th percentile of the $\mathcal{N}(0,T)$ distribution ($T = 184$), the 2 curves almost coincide at all dates. Figure \ref{f1_F0F2100} below gives a snapshot of this fact.

\begin{figure}[ht]\centering
	\includegraphics[width=\textwidth]{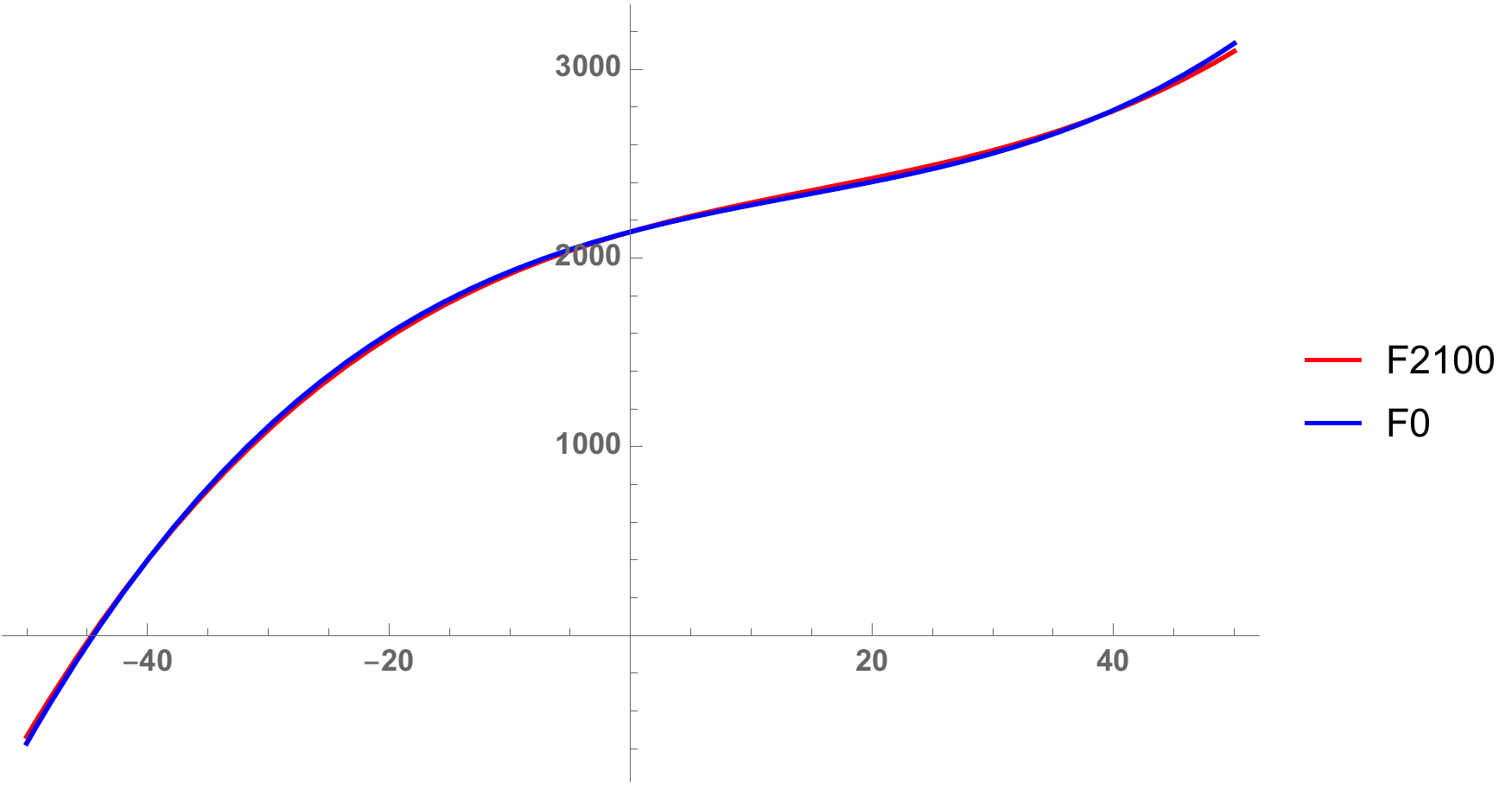}
	\caption{Representation of the functions retrieved using M1 ($F0$) and M2 ($F2100$)}
	\label{f1_F0F2100}
\end{figure}

As outlined earlier, the CMMV function $f_t(.)$ at every date is completely characterized by its expression at the final date $f_T(.)$. In figure \ref{f1_F0F2100}, the blue curve represent the polynomial approximation of $f_T$ ($F0$) retrieved using M1 (data on the prices of multiple options at date $t = 0$). The red curve represents the approximation of $f_T$ ($F2100$) retrieved using M2 (price data on the underlying and one option over 2 months). Hence, despite making use of 2 different types of data (cross-section for M1 and longitudinal for M2), we end up retrieving almost the same function associated with the CMMV. The 2 methods converging is to some extent a validation of the CMMV hypothesis and makes for a very good presage on the prediction phase that is coming next. Additionally, it is worth noting that the option prices we are working with do present a smile for the considered maturity. This brings up one important feature of the CMMV model: it does not assume a constant volatility. That is also the reason why we opt for the SS method rather than the Black-Scholes-Merton model for the comparative analysis. Finally, the volatilities for the SS pricing model are obtained using calibration at date $t = 0$ (same price data as for M1).

\subsubsection{Comparative analysis}

One way to compare the prediction accuracies of our methods to that of the SS method is to make a scatter plot of the observed option prices and overlay it with the 3 pricing functions. This would however require a distinct graph for each date in our test period, which ranges 4 months. And it turns out that for most of these dates, the graphs of the 3 pricing functions are hardly distinguishable. Therefore, we will make use of predicted prices. In details, after retrieving the CMMV function and estimating a polynomial model for the smile\footnote{We first compute the implied volatility of every option at date $t = 0$ and associate this vector of values to their corresponding strike prices. Then we use this list to build a polynomial fit for the volatility function. The need to use a volatility function stems from the fact that the set of traded options is not fixed throughout the trading period we consider. A number of options that are not traded at $t = 0$ appear at later dates.} at $t = 0$, we dispose of a pricing function for each model. These give a deterministic expression for the price of every option at every date as a function of the underlying's price. For each model, the value of these expressions when the underlying's price is plugged in is what we refer to as predicted prices.

For illustrative purposes, we compute the predicted prices for all options at every trading date ($t = 0$ to $t = 184$) from the 3 pricing functions. Then for each option and at each date, we compute the absolute difference between the predicted price and the observed price for every model (prediction error). In figure \ref{PredErrT} below, we represent the average\footnote{The average here is taken with respect to the strikes available.} (absolute) prediction error at every trading date for the pricing functions derived using M1 (blue dots), M2 (red dots) and SS (brown dots).

\begin{figure}[ht]\centering
	\includegraphics[width=\textwidth]{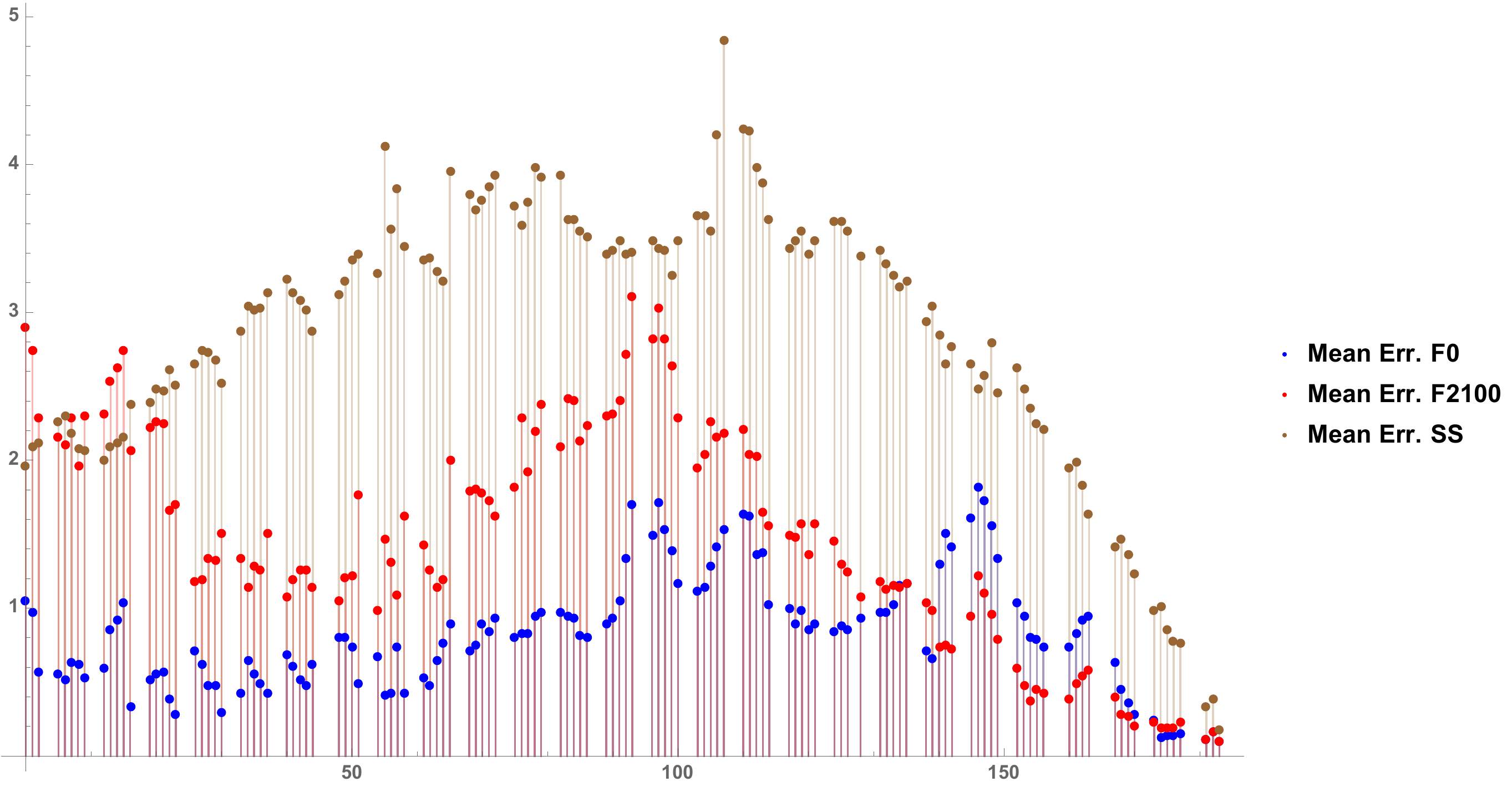}
	\caption{Evolution of the mean prediction errors over the trading period}
	\label{PredErrT}
\end{figure}

In light of the above figure, we observe that, bar a few days following the initial date, M1 and M2 outperform SS on average at every trading date. In fact, M1 makes consistently lower mean prediction errors. And unlike the 2 other methods (and what one could expect), its corresponding evolution of errors is somewhat closer to a horizontal line than it is to a bell-shaped curve. M2 catches up with it in terms of accuracy from the third month onwards, and even betters it on the fifth month. On top of that, the highest prediction error (the highest point on the graph) represents less than 2\% of the mean option price at its corresponding trading date. For M1, this relative mean error (mean absolute prediction error over the average observed price at each date) is less than 0.5\% throughout the 6 months. And for M2, the largest relative mean error is attained at period $t = 92$ and amounts to 0.8\%. From these, we can infer that the considered models do a good job pricing the options, especially considering the time span. Though these low mean prediction errors at each date might be the consequence of the low prices of out-of-the-money options. Thus, we next examine the mean prediction errors with respect to strikes.

\begin{figure}[ht]\centering
	\includegraphics[width=\textwidth]{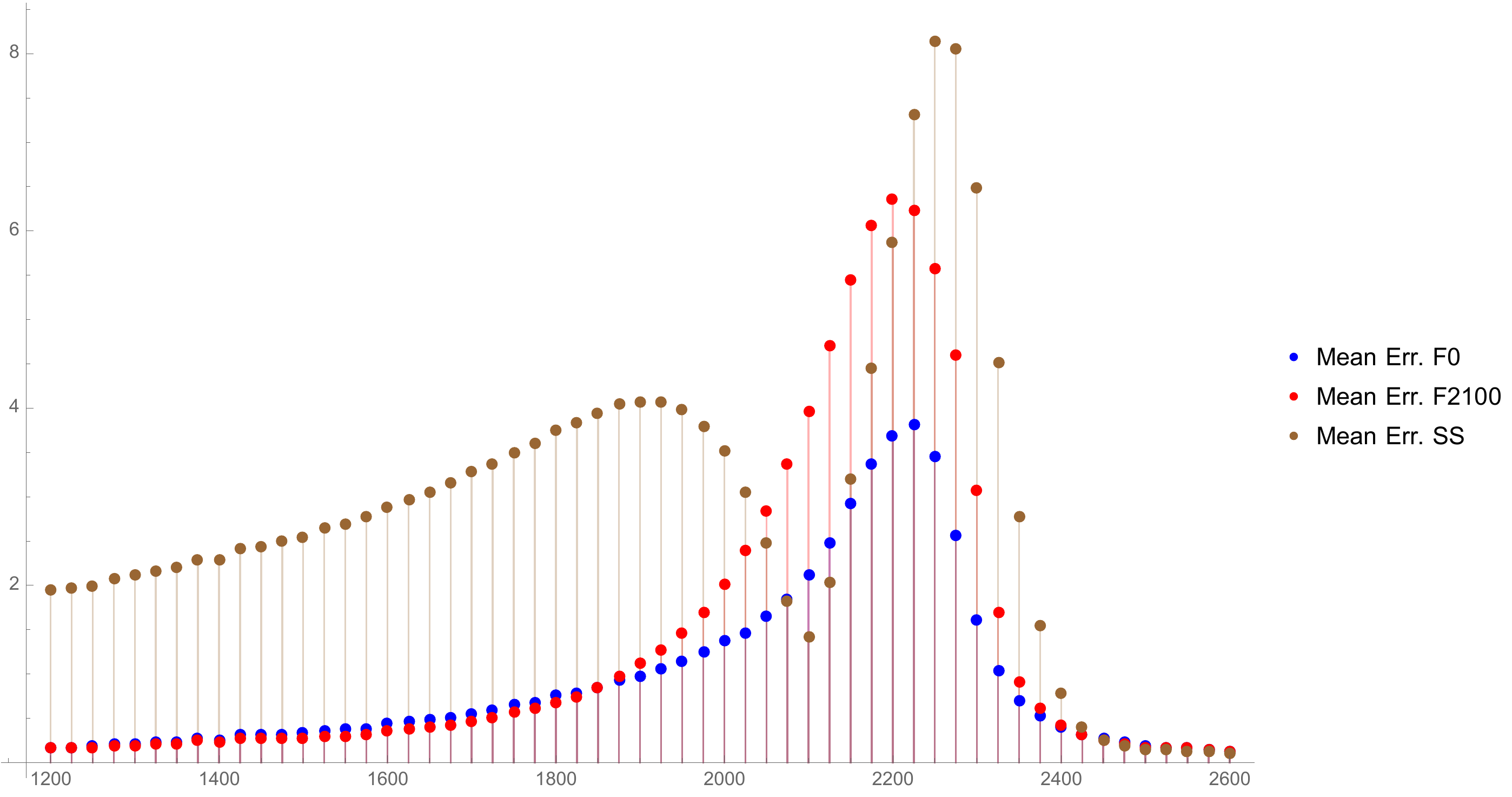}
	\caption{Representation of mean prediction errors with respect to strikes}
	\label{PredErrK}
\end{figure}

In this case, we again take the predicted prices for all options at every trading date and for each model. But instead of computing the absolute mean prediction error by trading date, we compute it by strike price. More succinctly, for every strike price, we determine the mean prediction error on the whole trading period, for each model. But we dispose of a large number of strikes at any trading date and not all options are traded at all dates. This is why we focus only on the most liquid options and the time frame of the test set\footnote{As mentioned earlier, the test set consists of the trading periods whose data was not used in the implementation of the function approximations. It spans the last 4 months (from $t = 62$ to $t = 184$) of the 6-months period we consider.} to reduce the computational cost. Figure \ref{PredErrK} below represents the absolute mean prediction errors for strikes ranging from 1200 to 2600, whose corresponding options are traded at every date in the test set.  

From figure \ref{PredErrK}, it appears that M1 and M2 actually outperform SS for almost all options in-the-money (the average spot price of the underlying over the period is 2170). SS does better than the CMMV models only on a small interval around the strike 2100. And this interval corresponds to the one on which the overlapping structure of the pricing functions' curves change: the curve of the SS pricing function moves above the 2 others. The relative errors (obtained by dividing the absolute mean error by the mean price for every strike), are less than 5\% for in-the-money options for all 3 models (less than 2\% for M1 and M2). Though, they attain quite large values for out-of-the-money options, which can be attributed to their low prices. One feature that is not perceptible from the graph of the absolute mean prediction errors is that for most of the trading dates, all 3 models underprice options with strikes lower than 2000 or larger than 2400. And options with strikes inside this interval are overpriced. 

All in all, figure \ref{PredErrK} reinforces the conclusion from figure \ref{PredErrT} about the precision of the CMMV models. The accuracy of M2 (or equivalently its pricing function being very close to that of M1) especially is startling, given the fact that only price data of a single option was used in its estimation. Not to mention its test set is twice as large as its training sample. And from the complementary analysis which is not presented in this paper, this is regardless of the chosen strike price, as long as it is liquid enough (in an interval ranging 500 around the average spot price of the underlying). 

This overall prediction performance of the CMMV models spurs the need for furthering the analysis. In fact after estimating the CMMV function for some underlying using M1 (or M2 for that matter), we can in theory, determine the price dynamics of any derivative on it, regardless of its type. Hence, in the following paragraphs, we explore to what extent this holds in practice, or one could say the \textit{robustness} of the models. 

\subsubsection{Further validation}

First, we consider the volatility surface. From the same initial date as in the previous paragraphs, the price dynamics derived using M1 with one maturity ($T = 184$) are utilized to determine the implied volatilities for 8 maturities in a one-year interval. In details, we first retrieve the CMMV function using M1 and price data of the options expiring at $T = 184$. Making use of the resulting pricing function, we compute the predicted prices of all options, for every maturity in the set: $\{30, 58, 93, 121, 149, 184, 240, 331\}$, then compute their corresponding impied volatilities (predicted implied volatilities). Since we also have options' (observed) price data for all these maturities in the dataset, we compute the observed implied volatilities by calibration with a classic Black-Scholes-Merton model. In figure \ref{f7_PredVS} below, we represent in a 3D plot, for every strike and every maturity, the observed (green dots) and predicted (blue dots) volatilities. 

\begin{figure}[ht]\centering
	\includegraphics[width=\textwidth]{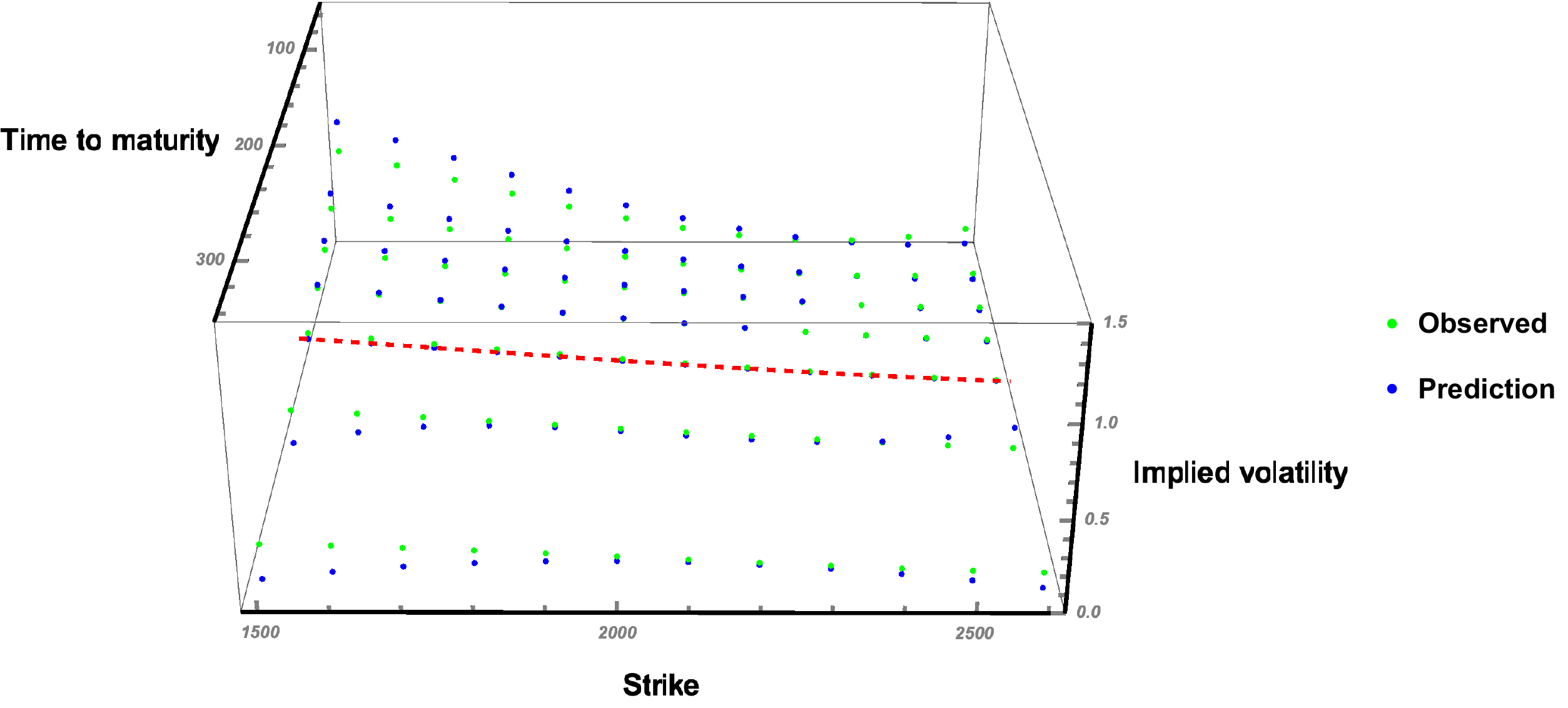} 
	\caption{Observed vs Predicted volatilities}
	\label{f7_PredVS}
\end{figure}
The red line indicates the implied volatilities corresponding to the maturity used for the implementation of M1 ($T = 184$). We observe from the graph that the predicted volatilities are very close to the observed ones. Indeed, around the spot price of the underlying at the initial date (2102.60), the predictions match the observed volatilities up to the hundredth decimal point for maturities greater than 2 months (58 days). Unsurprisingly, the gap between predictions and observations widens a bit as we drift away from this strike range. This can be explained by essentially 2 factors. First the options with strikes outside that range are typically less liquid. Hence, fewer data is available for the estimation of their prices. Secondly, the responsiveness of the implied volatility function is higher for these prices: a small variation in the prices can lead to a large gap in volatilities for options far in(or out-of)-the money. Nevertheless, the accuracy of the predictions for the strikes displayed on the graph suggests an impressive consistency of the CMMV function.

Still, practitioners argue that regarding the volatility surface, it is the short-term predictions that are particularly challenging. In order to assess the predictive performance of our model in the short run, we retrieve the CMMV function (with M1) using price data of options with the same exercise date as previously. However, we fix the initial date $t = 0$  just 10 days away from the expiry of options for which we make price predictions. Pointedly, we consider price data of all traded options (with expiry date January 20th) on December 6th, 2016 (maturity $T = 45$ days). Making use of M1, we derive an approximation of the CMMV function. The corresponding pricing function is then used to predict prices of options expiring on December 16th, 2016. In contrast to the previous illustrations, we consider relative average prediction errors (at each quotedate, the absolute mean prediction error across strikes is divided by the corresponding average option price) to emphasize the accuracy. They are represented on the graph below in which we compare our results to that of a SS model whose smile is fit using the same price data.

\begin{figure}[ht]\centering
	\includegraphics[width=\textwidth]{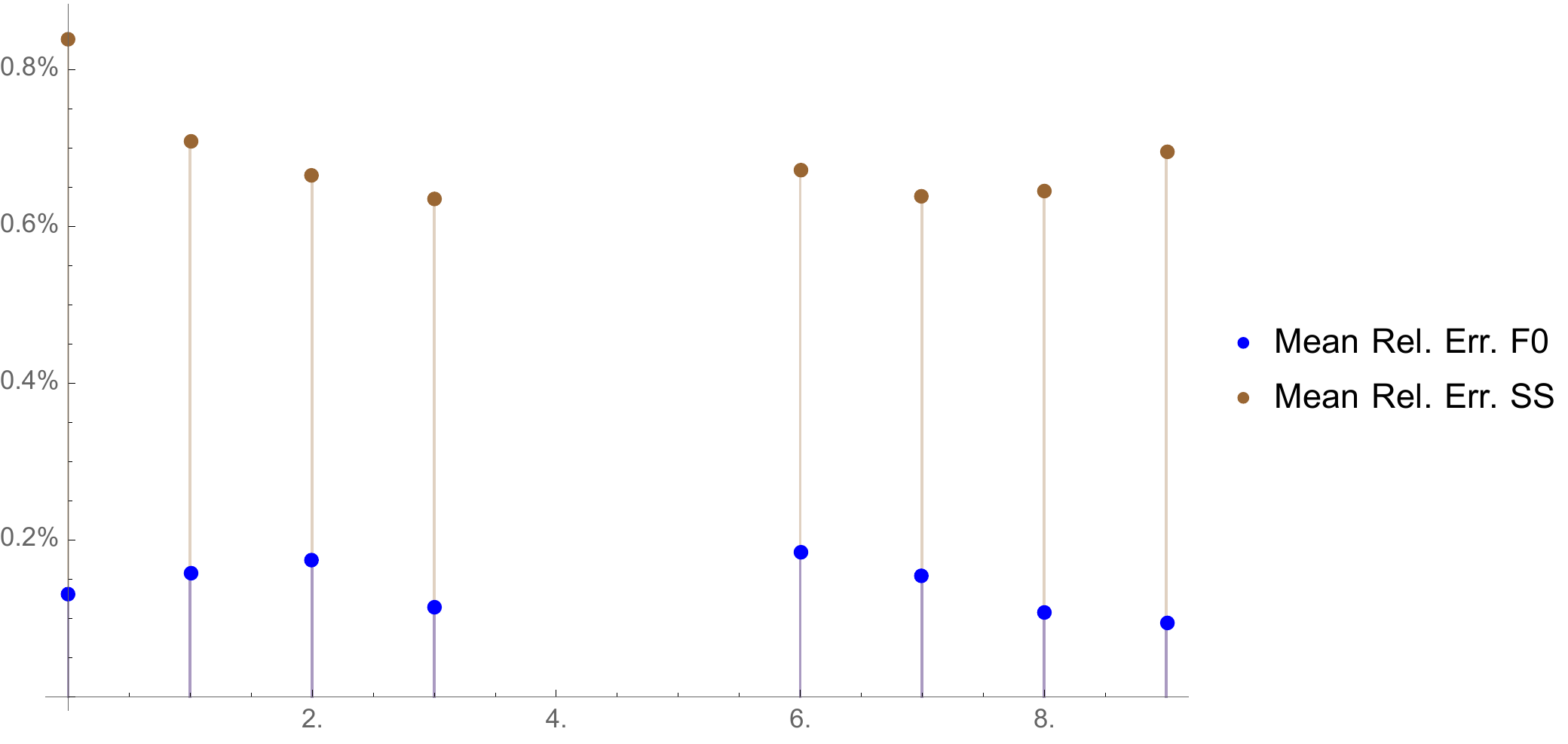}
	\caption{Evolution of relative mean prediction errors}
	\label{f8_Pred2T}
\end{figure}

From figure \ref{f8_Pred2T}, we observe that M1 edges the SS model once again. Though the key takaway here is the good predictive performance, despite the short maturity in consideration: the highest relative error is lower the 0.2\%. Actually, the same analysis was conducted with initial dates 15 and 4 days prior the second expiry date. In both cases, the results are practically identical to the ones presented on the above graph. We finish this section by providing our take on the comparison between local and stochastic volatility models.

\subsection{Local vs. Stochastic volatility models}

As mentioned in the description of the CMMV class, our model is a particular case of local volatility models. These were initially introduced by Dupire in \cite{dupire1994pricing}. In the model he introduces, the local volatility is recovered from the observed volatility surface at $t = 0$. Hence, to use this approach in practice, one needs the full knowledge of the volatility surface. This means that option prices are to be observed for all strikes and maturities. Unfortunately, for most securities, there are usually very few options with different maturities available on the market in a given year. This is why Dupire's method necessitates interpolation of the volatility surface and its results can be very sensitive to the selected class of interpolating functions. In contrast M1 only requires the observations of option prices for all strikes for a single maturity. And for M2, one only needs the price process of one derivative in addition to the process of the underlying to determine the price dynamic. The availability of the required data makes the CMMV models more appealing than the more general local volatility model. Note that in theory, Dupire's method applied to a CMMV would give the same price predictions but would need much more data to get the same accuracy. 

Either way, local volatility models get a fair share of criticism, especially regarding the approximation of the volatility surface. In particular, they are alleged to fail to give appropriate smile dynamics. For instance Hagan et al. (2002) \cite{hagan2002managing} suggest that, the smile curve from the local volatility models moves in opposite direction to the spot price of the underlying, which contradicts the market. This is stated as an argument for the use of stochastic volatility models. It was therefore an interesting point to check whether this is the case for our model. Consequently, we fit the CMMV function (with M1) using 3 different spot prices for the underlying: the actual spot price and 2 values either side of it. In the graphs below, we represent the corresponding implied volatilities (from the pricing function with the actual spot price in blue) and it turns out our smile curve does move in the right direction.

\begin{figure}[ht]\centering
	\begin{minipage}[b]{.48\textwidth}
		\subfloat[price change from 2138 to 2038]{{\includegraphics[width=\textwidth]{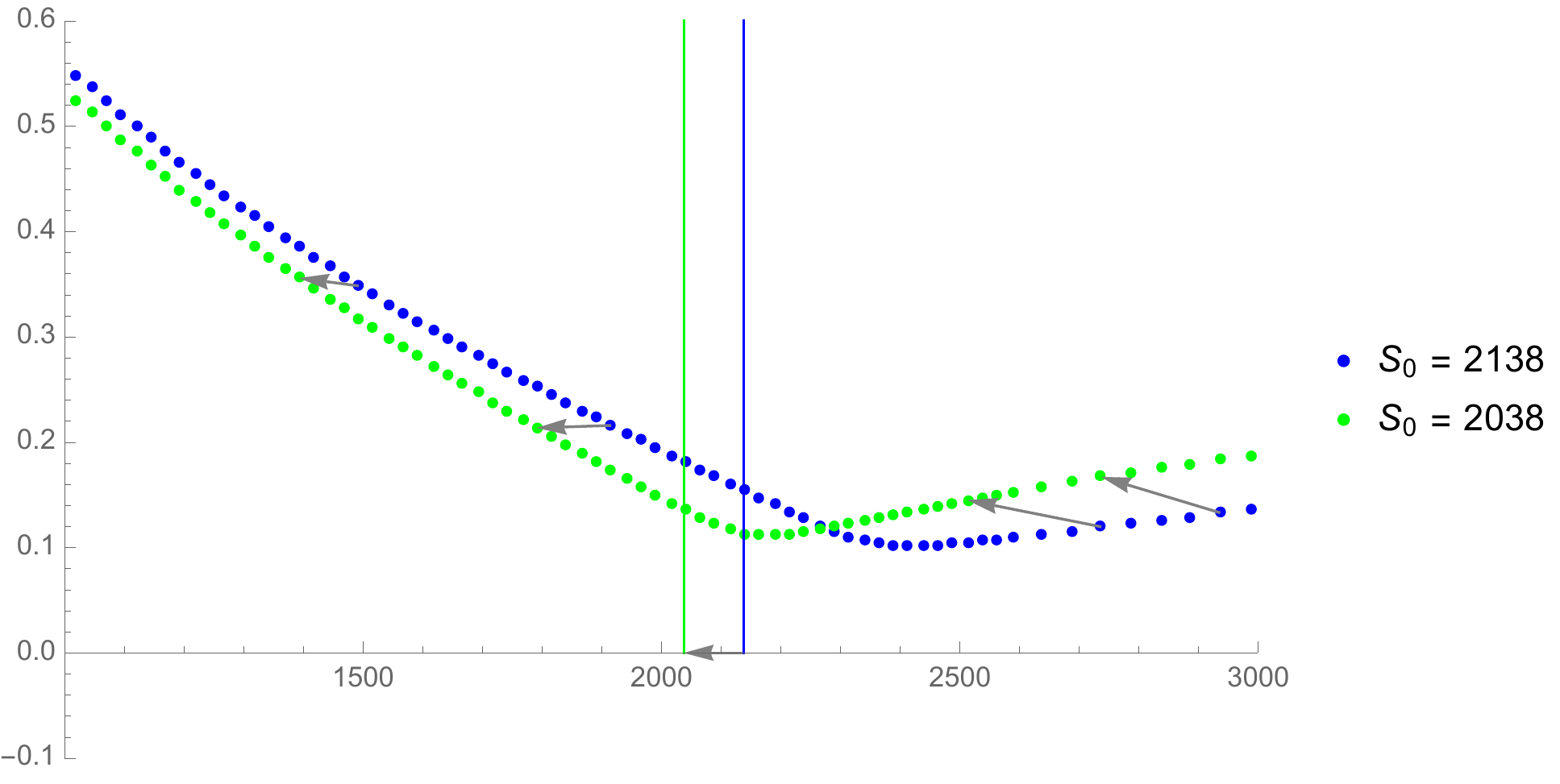} }}%
	\end{minipage}
	\quad
	\begin{minipage}[b]{.48\textwidth}
		\subfloat[price change from 2138 to 2200] {{\includegraphics[width=\textwidth]{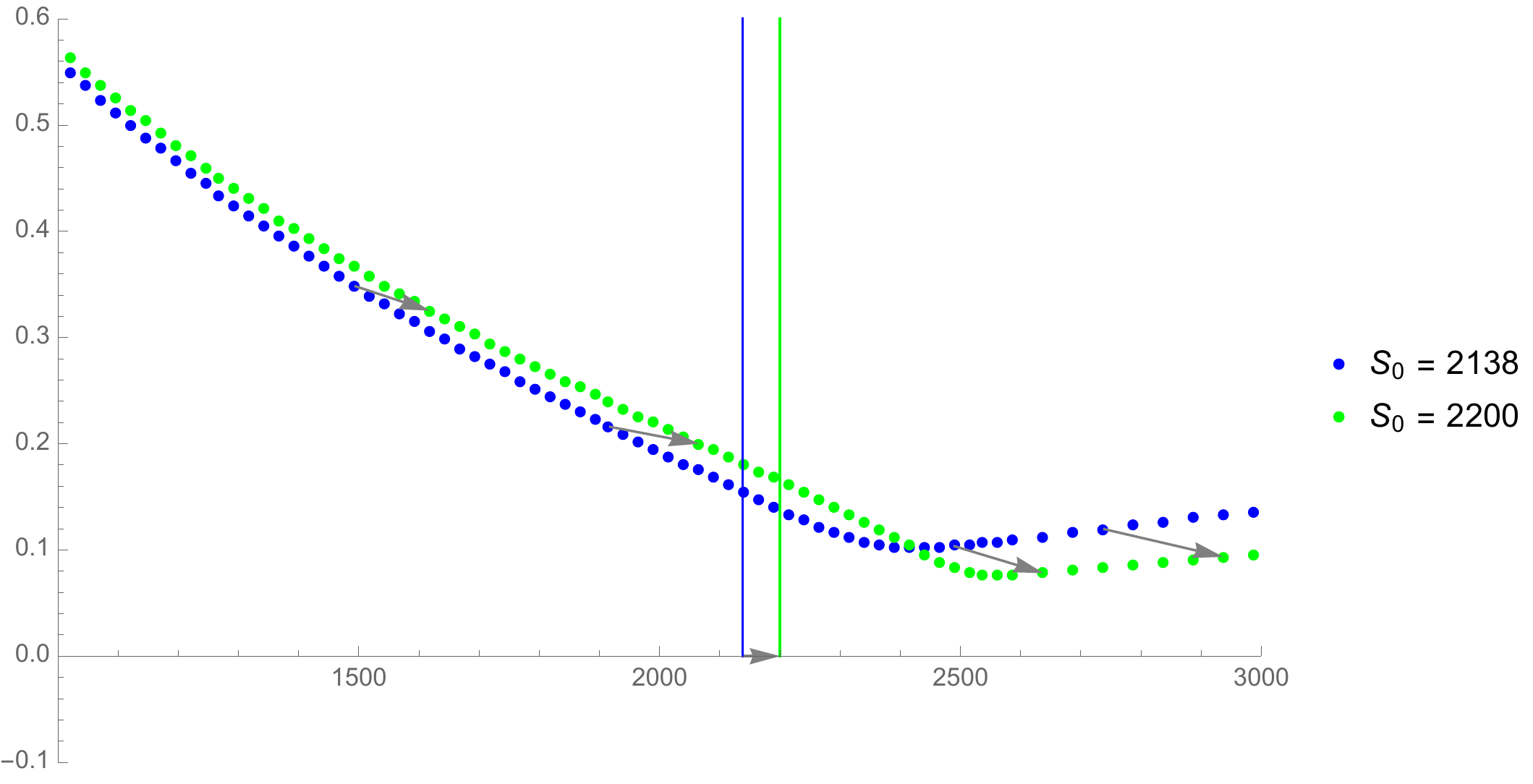} }}%
	\end{minipage}
	~\\[.3cm]
	\caption{Representation of the smile shift resulting from spot price movements}
	\label{SmileShift}
\end{figure}

The vertical lines correspond to the underlying's spot prices and these graphs provide evidence that not all local volatility models give wrong smile dynamics. Actually, a closer look at Hagan et al.'s proof exibits the fact that they assume a piecewise time-independent local volatility function. This is definitely not the case for the CMMV model. 

We do not make a direct comparison between our model and a stochastic volatility one in this paper because just like Dupire's model, they require more data. This is, nevertheless, an interesting question to explore in an attempt to gauge the CMMV model. In any case, the local volatility models seem to do a better job fitting the volatility surface than their critics suggest. For example, Podolskij and Rosenbaum \cite{podolskij2012testing} test the local volatility model assumption for the price dynamics against the alternative of a stochastic volatility model using historical prices. Their main conclusion is that there is no statistical evidence in favor of rejecting the local volatility assumption.

\section*{Conclusion}

The aim of this paper is twofold: present the CMMV model and give some evidence about its ability to fit market data. Thus, its market microstructure roots and most important features are highlighted in the first 2 sections. Next, 2 option pricing methods based on its paradigm are presented and used to fit historical vanilla option price data. These 2 methods, although requiring different types of data (cross section for the first and longitudinal for the second), give very similar and accurate results. Furthermore, the CMMV model has a very attractive consistency property: not only the price of the underlying is a CMMVs, but also the prices of European call and put options belong to the same class of dynamics.

All in all, the numerical results presented in this paper make a good case for the CMMV model and call for further assessment. In particular, it would be interesting to see how it fares in pricing exotic options and/or in comparison to models that theoretically would make better fits (and require more data).

\section*{acknowledgements}
The authors would like to hereby express their gratitude to O. Guéant, E. Abi Jaber, L. Campi, C. Martini for very insightful comments and suggestions.

\printendnotes

\bibliography{Bibliography}

\begin{thebibliography}{14}
\providecommand{\natexlab}[1]{#1}
\providecommand{\url}[1]{\texttt{#1}}
\providecommand{\urlprefix}{}

\bibitem[{Almgren and Chriss(2001)Almgren, Robert and Chriss,
  Neil}]{almgren2001optimal}
Almgren R, Chriss N.
\newblock Optimal execution of portfolio transactions.
\newblock Journal of Risk 2001;3:5--40.

\bibitem[{De~Meyer and Saley(2003)De Meyer, Bernard and Saley, Hadiza
  Moussa}]{meyer2003strategic}
De~Meyer B, Saley HM.
\newblock On the strategic origin of Brownian motion in finance.
\newblock International Journal of Game Theory 2003;31(2):285--319.

\bibitem[{De~Meyer(2010)De Meyer, Bernard}]{de2010price}
De~Meyer B.
\newblock Price dynamics on a stock market with asymmetric information.
\newblock Games and economic behavior 2010;69(1):42--71.

\bibitem[{Gensbittel(2010)Gensbittel, Fabien}]{gensbittel2010analyse}
Gensbittel F.
\newblock Analyse asymptotique de jeux r{\'e}p{\'e}t{\'e}s {\`a} information
  incompl{\`e}te.
\newblock PhD thesis, Universit{\'e} Panth{\'e}on-Sorbonne-Paris I; 2010.

\bibitem[{De~Meyer and Fournier(2017)De Meyer, Bernard and Fournier,
  Ga{\"e}tan}]{de2017price}
De~Meyer B, Fournier G.
\newblock Price dynamics on a risk-averse market with asymmetric information.
\newblock arXiv preprint arXiv:170103341 2017;.

\bibitem[{Dupire et~al.(1994)Dupire, Bruno and others}]{dupire1994pricing}
Dupire B, et~al.
\newblock Pricing with a smile.
\newblock Risk 1994;7(1):18--20.

\bibitem[{Beck et~al.(2004)Beck, Vincent and Malick, J{\'e}r{\^o}me and
  Peyr{\'e}, Gabriel}]{beck2004objectif}
Beck V, Malick J, Peyr{\'e} G.
\newblock Objectif agr{\'e}gation.
\newblock H\&K; 2004.

\bibitem[{Revuz and Yor(1999)Revuz, Daniel and Yor, Marc}]{revuz1999continuous}
Revuz D, Yor M, Continuous Martingales and Brownian Motion, vol. 293 of
  Fundamental Principles of Mathematical Sciences.
\newblock Springer, Berlin, Germany,; 1999.

\bibitem[{Akimoto et~al.(2012)Akimoto, Youhei and Auger, Anne and Hansen,
  Nikolaus}]{akimoto2012convergence}
Akimoto Y, Auger A, Hansen N.
\newblock Convergence of the Continuous Time Trajectories of Isotropic
  Evolution Strategies on Monotonic C 2$\backslash$mathcal C\^{} 2-composite
  Functions.
\newblock Parallel Problem Solving from Nature-PPSN XII 2012;p. 42--51.

\bibitem[{Hansen(2016)Hansen, Nikolaus}]{hansen2016cma}
Hansen N.
\newblock The CMA evolution strategy: A tutorial.
\newblock arXiv preprint arXiv:160400772 2016;.

\bibitem[{Hansen(2006)Hansen, Nikolaus}]{hansen2006cma}
Hansen N.
\newblock The CMA evolution strategy: a comparing review.
\newblock Towards a new evolutionary computation 2006;p. 75--102.

\bibitem[{Peng et~al.(2010)Peng, Fei and Tang, Ke and Chen, Guoliang and Yao,
  Xin}]{peng2010population}
Peng F, Tang K, Chen G, Yao X.
\newblock Population-based algorithm portfolios for numerical optimization.
\newblock IEEE Transactions on Evolutionary Computation 2010;14(5):782--800.

\bibitem[{Hagan et~al.(2002)Hagan, Patrick S and Kumar, Deep and Lesniewski,
  Andrew S and Woodward, Diana E}]{hagan2002managing}
Hagan PS, Kumar D, Lesniewski AS, Woodward DE.
\newblock Managing smile risk.
\newblock The Best of Wilmott 2002;1:249--296.

\bibitem[{Podolskij and Rosenbaum(2012)Podolskij, Mark and Rosenbaum,
  Mathieu}]{podolskij2012testing}
Podolskij M, Rosenbaum M.
\newblock Testing the local volatility assumption: a statistical approach.
\newblock Annals of Finance 2012;8(1):31--48.

\end{thebibliography}

\end{document}